\documentclass[11pt]{article}

\newcommand{\Bz}{\ensuremath{\langle B_z\rangle}}
\def\gtrsim{\mathrel{\hbox{\rlap{\hbox{\lower4pt\hbox{$\sim$}}}\hbox{$>$}}}}

\def\ltsim{\mathrel{\hbox{\rlap{\hbox{\lower4pt\hbox{$\sim$}}}\hbox{$<$}}}}

\def\kms{\hbox{{\rm km}\,{\rm s}$^{-1}$}}

\usepackage[]{sao1}
\usepackage{graphicx}
\begin{document}

\title{Magnetic fields in Herbig Ae/Be stars}
\author{G.A. Wade$^1$, C. Catala$^2$,  E. Alecian$^2$, C. Folsom$^{1,3}$, S. Bagnulo$^4$, T. B\"ohm$^5$, J.-C. Bouret$^6$, J.-F. Donati$^5$, D. Drouin$^1$, J.D. Landstreet$^7$, }
\institute{$^1$Department of Physics, Royal Military College of Canada, Kingston, Canada  
\and 
   $^2$Obs. de Paris LESIA, Meudon, France
\and 
   $^3$Department of Physics, Queen's University, Kingston, Canada
\and 
   $^4$European Southern Observatory, Santiago, Chile
\and 
   $^5$Obs. Midi-Pyr\'en\'ees, Toulouse, France
\and 
   $^6$Laboratoire d'Astrophysique de Marseille, Marseille, France
\and 
   $^7$Department of Physics \& Astronomy, University of Western Ontario, London, Canada}

\maketitle

\begin{abstract}

Studies of stellar magnetism at the pre-main sequence phase can provide important new insights into the detailed physics of the late stages of star formation, and into the observed properties of main sequence stars. This is especially true at intermediate stellar masses, where magnetic fields are strong and globally organised, and therefore most amenable to direct study. This paper reviews recent high-precision spectropolarimetric observations of pre-main sequence Herbig Ae/Be stars, which are yielding qualitatively new information about intermediate-mass stars: the origin and evolution of their magnetic fields; the role of magnetic fields in generating their spectroscopic activity and in mediating accretion in their late formative stages; the factors influencing their rotational angular momentum; and the development and evolution of chemical peculiarity in their photospheres.

\end{abstract}

\section{Introduction}

Herbig Ae/Be (HAeBe) stars are pre-main sequence stars of intermediate mass (Herbig 1960; Hillebrand et al. 1992). They are characterised by spectral types
A and B with strong, often ubiquitous emission lines. They are distinguished
from the classical Ae/Be stars by their IR colours and frequent
presence within dust-obscured regions and association with nebulae
(Waters \& Waelkens 1998). According to stellar evolution theory, HAeBe stars should not posses {deep} outer convection zones {which generate the important quantities of outward-flowing mechanical energy required to power an MHD dynamo.} Rather, these stars are expected to have convective cores surrounded by {primarily} radiative sub-photospheric envelopes (Iben 1965; Gilliland 1986). However, since 1980, repeated observations (e.g. Praderie et al. 1982; Catala et al. 1986;
Hamann \& Persson 1992; Pogodin et al. 2005) have shown that many HAeBe stars are intensely active. In particular, some stars display characteristics often associated with magnetic activity and the presence of chromospheres or coronae (e.g. Skinner \& Yamauchi 1996; Hamaguchi et al. 2000). These properties {have been proposed as indicators that} many of these stars {or their circumstellar envelopes} are intensely magnetically active. 

This proposal has important implications for our picture of how intermediate-mass stars form. In lower-mass pre-main sequence T Tauri stars, it is now generally supposed that accretion is mediated by the presence of strong, large-scale magnetic fields (e.g. discussion by Johns-Krull et al. (1999) and references therein). Some authors (e.g. Muzzerole et al. 2004) have suggested that similar ``magnetospheric accretion'' may occur in intermediate-mass PMS stars as well. 

Insight into the magnetic properties of HAeBe stars can likely be gleaned from the magnetic properties of their better-studied main sequence descendants. A few percent (e.g. Wolff 1968) of all A and B type {main sequence stars} exhibit organised magnetic fields with strengths ranging from a few
hundred to a few tens of thousands of gauss (e.g. Borra \& Landstreet 1980; Mathys et al. 1997; Bagnulo et al. 1999). Because, like their pre-main sequence progenitors, main sequence A and B type stars have radiative envelopes, these magnetic fields are not believed to result from a contemporaneous MHD dynamo. Rather, they are thought to be {\em fossil fields}: the passively-decaying remnants of magnetic fields produced at an earlier convective evolutionary stage, or swept up during star formation. The presence of these fields has important consequences for the structure of
the atmospheres of these {\em Ap/Bp stars}, suppressing {large-scale mixing} and
leading to strong atmospheric chemical peculiarities and abundance patches (e.g. Folsom et al. 2006; Lueftinger et al. 2003). Although the separation and mixing processes leading to these chemical peculiarities and abundance structures are not well understood, the peculiarities are sufficiently unique to the Ap/Bp that they allow for the robust identification of magnetic stars from high- and even moderate-resolution spectroscopy (i.e. without any direct detection of the magnetic field). The presence of the field also appears to strongly influence stellar rotation, with Ap/Bp stars rotating substantially more slowly than other stars of the same spectral type (e.g. Abt \& Morrell 1995). As proposed by St\c{e}pie\'n (2000), the slow rotation of Ap/Bp stars can be explained by angular momentum loss at the pre-main sequence phase, taking into account accretion of matter along the magnetic field lines, the stellar field-disc interaction and a magnetised stellar wind.

On the other hand, high-precision, high-resolution spectropolarimetric studies (e.g. Shorlin et al. 2002, Wade et al. 2006a) have demonstrated conclusively that the large majority of main sequence A and B type stars (i.e. essentially all non-Ap/Bp stars) show no direct or indirect evidence for magnetic fields, often with remarkably small upper limits (a few gauss). The existence of this ``magnetic dichotomy'' (i.e. the qualitative difference between the magnetic properties of the Ap/Bp stars and the other main sequence A and B stars) is further reinforced by the results of Auri\`ere et al. (in preparation, but see Auri\`ere et al. 2004), who show convincingly that all spectroscopically-identified Ap/Bp stars indeed host magnetic fields which are detectable with current techniques, and which are characterised by surface dipole components stronger than about 200~G. Although these magnetic properties of A and B type main sequence stars are not understood in theoretical detail, they are now well-established, providing solid observational results with which to compare their pre-main sequence progenitors.

Observations of magnetic fields in Herbig Ae/Be stars can therefore serve to address several important astrophysical problems: (1) The role of magnetic field in mediating accretion, and the validity of models which propose that HAeBe stars are simply higher-mass analogues of the T Tau stars. (2) The origin and properties of the magnetic fields of main sequence A and B type stars. (3) The development and evolution of chemical peculiarities and chemical abundance structures in the atmospheres of A and B type stars. (4) The loss of rotational angular momentum which leads to the slow rotation observed in main sequence A and B type stars.

This paper will review recent magnetic observations of Herbig Ae/Be stars, discussing how these new data help to clarify the issues listed above.

\section{Low-resolution spectropolarimetry with FORS1 at the ESO-VLT}

Wade et al. (2006b) presented results of a survey of the longitudinal magnetic fields of HAeBe stars conducted using the low-resolution FORS1 spectropolarimeter at the VLT observatory. This work is the largest and most sensitive search for magnetic fields in HAeBe stars published to date. 

\begin{figure*}

\centering

\includegraphics[width=15cm]{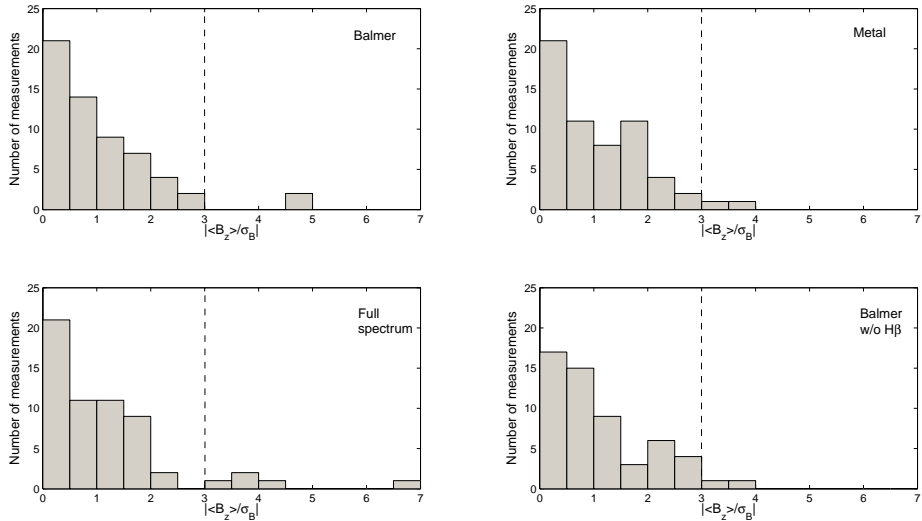}

\caption{Histograms of the detection significance $z=\Bz/\sigma_B$ of magnetic field measurements of stars observed by Wade et al. (2006b). Each panel corresponds to one of the 4 measurements obtained using different parts of the spectrum. The dashed line indicates the $3\sigma$ detection threshold. Nearly all measurements correspond to non-detections, and formal detections are obtained for only 4 stars.}\label{zBz}

\end{figure*}

The primary goal of this survey was to determine whether strong, organised magnetic fields, similar to those of main sequence Ap/Bp stars, could be detected in the HAeBe stars, and if so, to determine the intensities, geometrical characteristics, and statistical incidence of such fields.

With this aim, Wade et al. (2006b) obtained 68 high-S/N ($\sim 3000:1$), low resolution (typically $R\sim 1000-1500$) observations of 50 HAeBe stars in circularly polarised light. To obtain a high signal-to-noise ratio and to mitigate spurious polarisation signals, they obtained multiple exposures of each star, with alternate exposures corresponding to a quarter-wave plate orientation of either $+45^{\rm{o}}$ or $-45^{\rm{o}}$. The data were reduced and analysed using the procedures of Bagnulo et al. (2006), developed and tested in the context of a survey of magnetic fields in 235 magnetic and non-magnetic A and B type stars in open clusters. 

The longitudinal magnetic field was inferred from the polarisation spectra using a linear least-squares fit based on the predicted circularly polarised flux $V/I$ in the weak-field regime, taking into account uncertainties associated with each reduced spectral pixel as described by Bagnulo et al. (2006). This procedure was repeated 4 times for each spectrum, diagnosing the field from all Balmer lines, all Balmer lines except H$\beta$, metallic lines, and the full spectrum. The mean 1$\sigma$ longitudinal field uncertainty measured from Balmer lines was 66~G, while that measured using the full spectrum was 48~G. 
 
The histograms of the detection significance $z=|\Bz/\sigma_B|$ of the resultant measurements are shown in Fig. 1. To robustly treat cases for which the longitudinal field was detected at about the 3$\sigma$ level, and in which minor changes in the data reduction would transform a marginal detection in a null or into a
definite detection, Wade et al. (2006b) evaluated the consistency of the longitudinal fields determined from the various spectral regions using two detection criteria described by Bagnulo et al. (2006). This allowed them to realistically judge the true significance of apparent detections.

\begin{figure*}

\centering

\includegraphics[width=9cm]{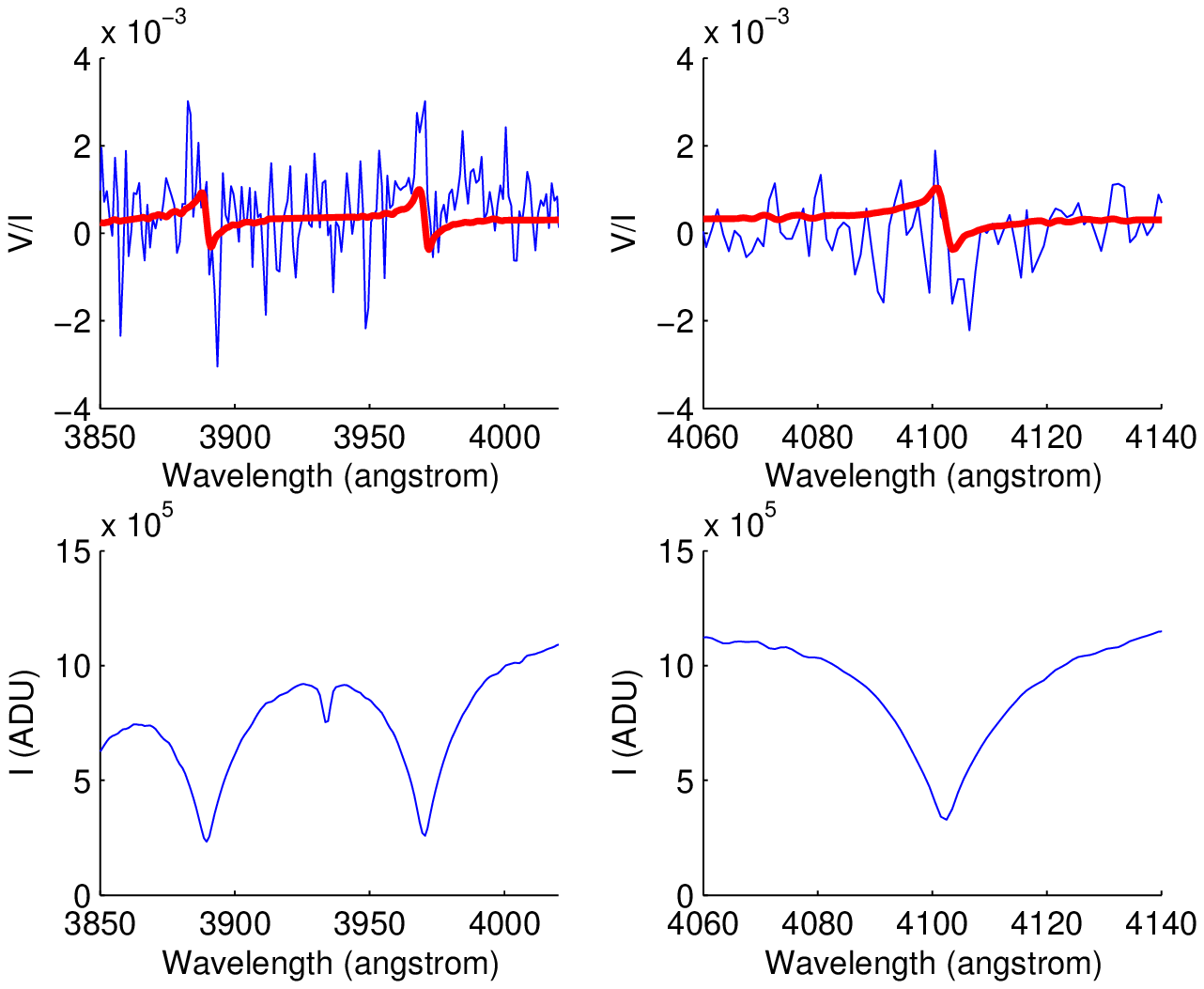}\includegraphics[width=9cm]{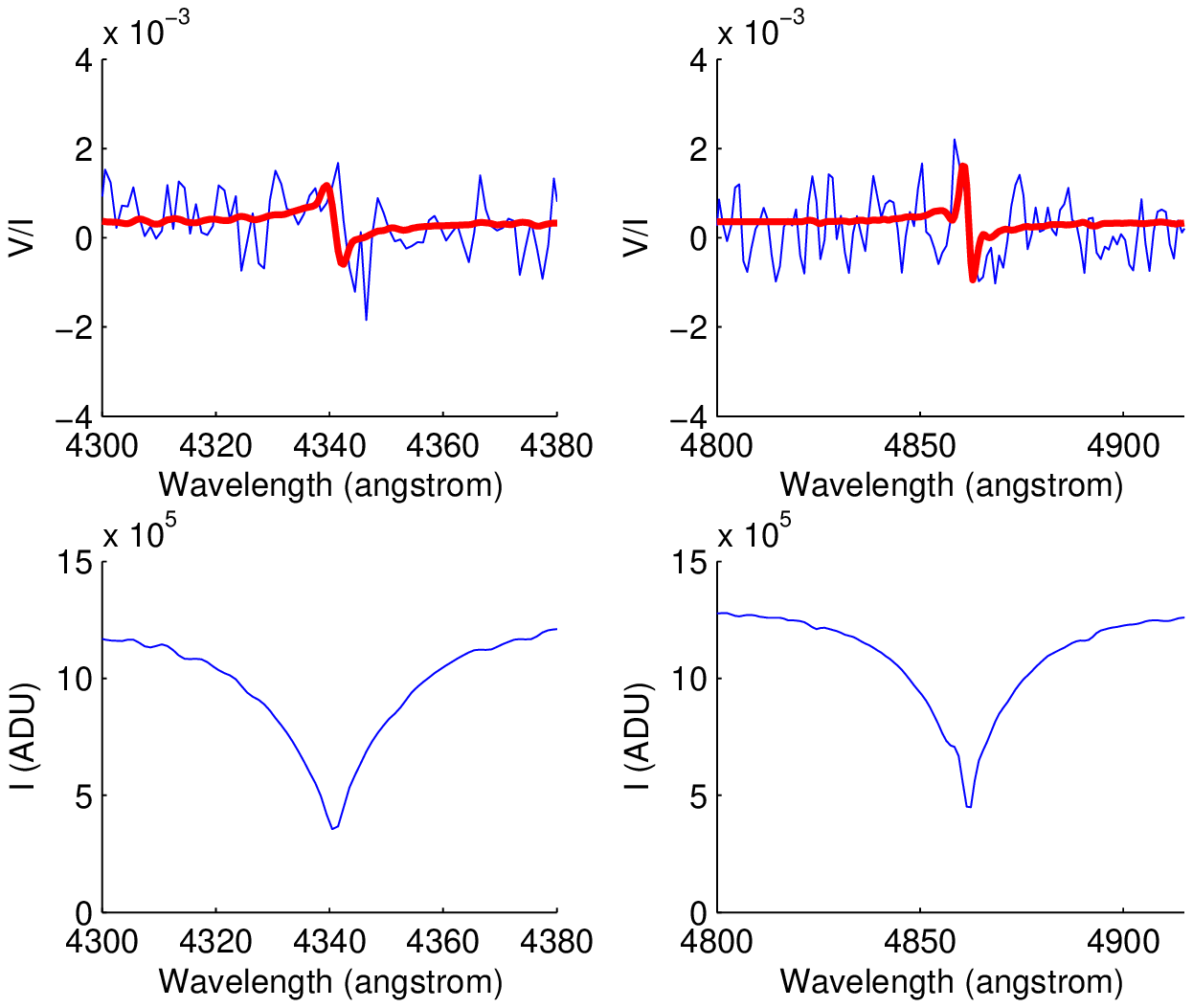}
\caption{Stokes $I$ and $V$ profiles in the FORS1 spectrum of HD 101412. The solid curve in the upper frames represents the prediction of the weak-field model. From Wade et al. (2006b).}

\end{figure*}

Ultimately, Wade et al. (2006b) concluded tentatively that weak longitudinal magnetic fields were detected in spectra of two HAeBe stars: HD 101412 (obtaining a measurement of $512\pm 111$~G from Balmer lines) and BF Ori (obtaining a measurements of $-180\pm 38$~G from Balmer lines). Selected regions of the spectrum of HD 101412 are shown in Fig. 2, overlaid with the model $V/I$ spectrum computed using the weak field model and the measured value of the longitudinal field. Although no obvious polarisation signatures are visible in individual absorption lines in Fig. 2, this does not necessarily indicate the absence of a field. As explained at the end of Sect.~8 and in Fig.~8 of Bagnulo et al. (2006), in their survey of cluster magnetic Ap stars they encountered
numerous cases where a simple inspection of the $V/I$ profiles did not show
any obvious polarisation signal, but in which the magnetic field was detected by
the linear regression of $V/I$. The reliability of this result is supported
by the histograms of Fig.~7 of Bagnulo et al. (2006), showing that virtually
no detections were found in "normal" A stars, where they do not expect to find
any magnetic field. Magnetic fields were detected only in Ap stars, i.e.,
where they are expected to be found. 

Magnetic observations of a small number of HAeBe stars have also been carried out using FORS1 by Hubrig et al. (2004, 2006a, 2006b), who reported marginal detections of magnetic fields in several stars. Of these, Wade et al. (2006b) observed HD~144432 and HD~31648, obtaining no detection in either. {Hubrig et al. (2006b) also reported the presence of strong Stokes $V$ signatures in spectral lines of some HAeBe stars, which they interpreted to be due to ``circumstellar'' magnetic fields. Wade et al. (2006b) were unable to confirm the detection of such features in their own spectra. As a further test, to check if they could confirm the strong polarisation in the same data employed by Hubrig et al., they extracted from the ESO Science Archive 3 spectra of the HAeBe star HD 190073 discussed by Hubrig et al. (2006b), reducing and analysing them according to the procedures of Bagnulo et al. (2006). Remarkably, Wade et al. found no evidence for the presence of strong, systematic polarisation signatures as reported by Hubrig et al. (2006b). This implies that the presence of this polarisation is highly sensitive to the details of the data reduction, suggesting that it is not astrophysical in origin.

\section{High-resolution spectropolarimetry with ESPaDOnS at the CFHT}

Wade, Catala and collaborators have also undertaken a large circular polarisation survey of HAeBe stars using the new high-resolution ESPaDOnS spectropolarimeter at the Canada-France-Hawaii Telescope. Although complete, the general results of this survey have not yet been published. However, first results and results for a few individual stars of great interest have been reported by Wade et al. (2005), Catala et al. (2006), Folsom et al. (these proceedings) and Alecian et al. (these proceedings).

\begin{figure*}
\centering
\includegraphics[width=6cm,angle=-90]{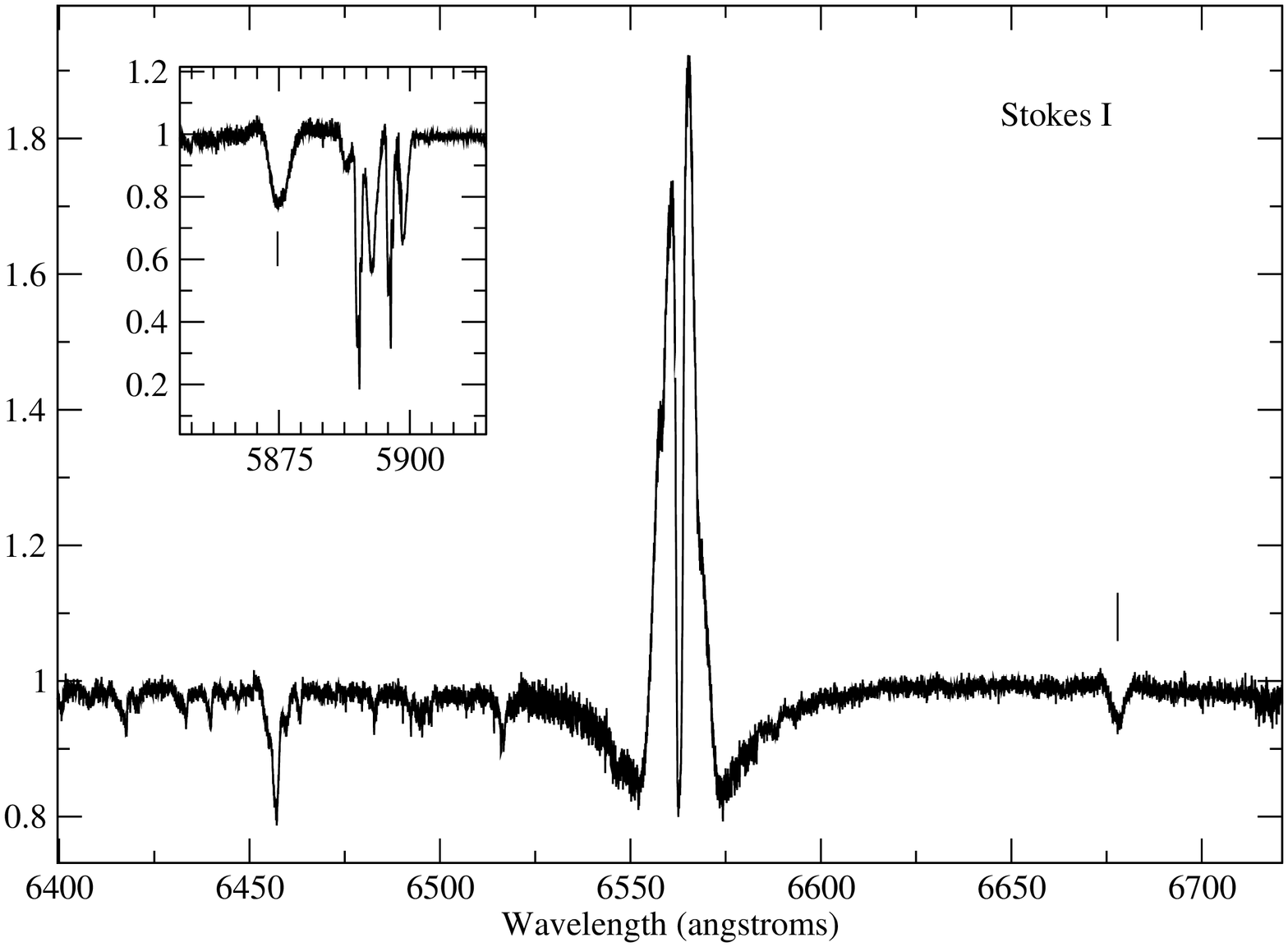}\includegraphics[width=6cm,angle=-90]{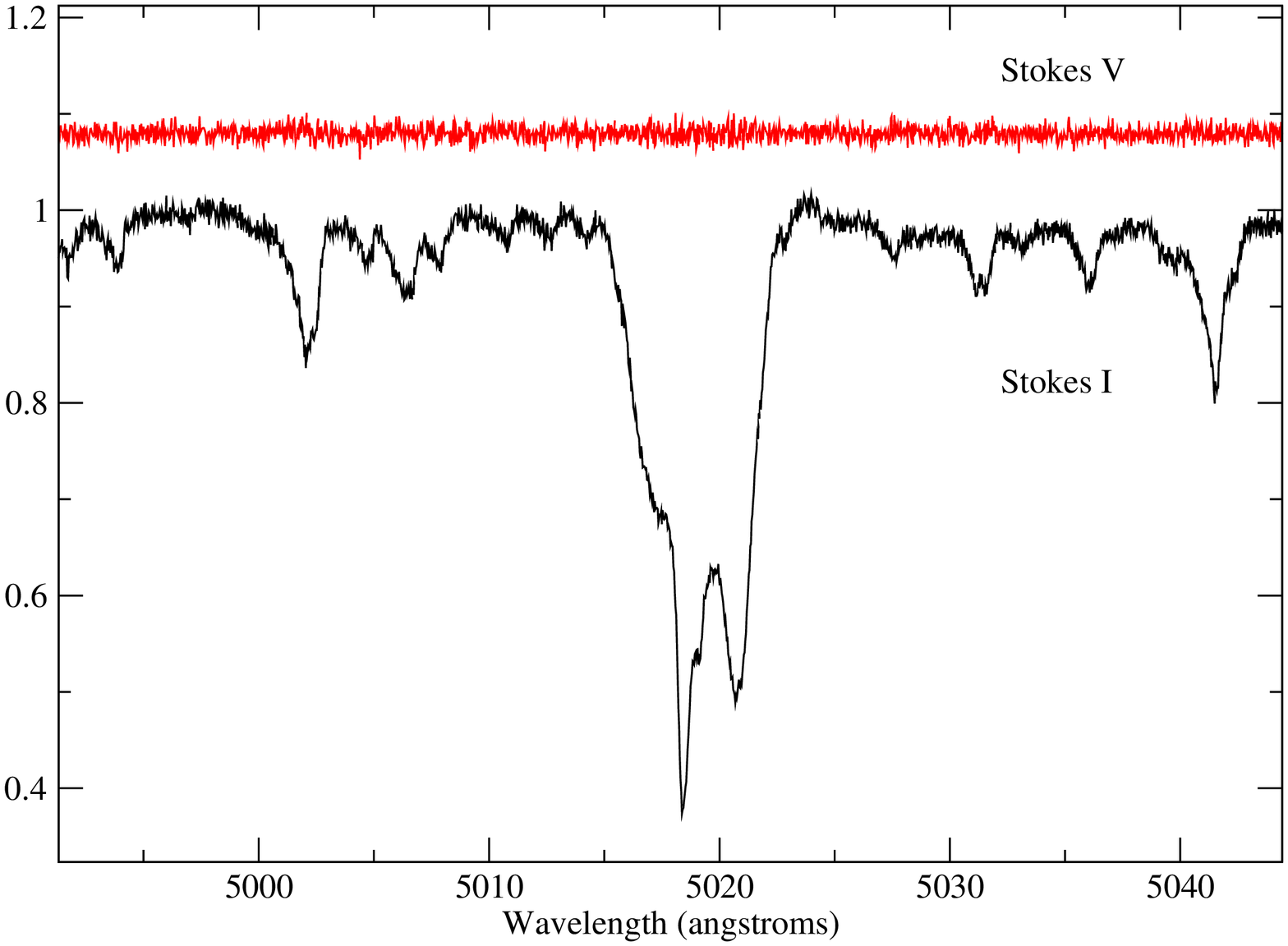}
\caption{{\em Left -}\ ESPaDOnS Stokes $I$ spectrum of the HAeBe star BF Ori, showing the emission profile of H$\alpha$. Absorption profiles of He~{\sc i}~$\lambda 6678$ and He~{\sc i}~$\lambda 5876$ (inset) are indicated by ticks. {\em Right -}\ Detail of the ESPaDOnS Stokes $I$ and $V$ spectra of BF~Ori showing the complex absorption profile of Fe~{\sc ii}~$\lambda 5018$. No circular polarisation is detected across this line.}
\end{figure*}

The ESPaDOnS survey comprises over 100 circular polarisation spectra of about 50 HAeBe stars. Some of the stars observed by Wade et al. (2006b) and Hubrig et al. (2004, 2006a,b) were also observed using ESPaDOnS. The spectra cover the range 3800-10400~\AA\ with a resolving power of approximately 65000, and a median peak S/N of about 300:1 per 2.6~\kms\ pixel (equivalent to a S/N of about 1450:1 per \AA\ at 5000~\AA). This is about one-half the mean S/N of the FORS1 spectra of Wade et al. (2006b); it furthermore varies more strongly with wavelength, decreasing to about 1/2 its peak value at 4000~\AA\ and 9000~\AA. However, the excellent results obtained with ESPaDOnS so far (see Donati et al. 2005, 2006ab; Wade et al. 2006; Catala et al. 2006, 2007) demonstrate that the advantage of substantially higher resolving power (45-65 times better resolution) easily compensates for this difference. In particular, the ESPaDOnS measurements are characterised by a substantially higher magnetic field detection rate, more robust detections of weak longitudinal fields, straightforward confirmation of field detections, a lower ultimate detection threshold, and sensitivity to more complex field topologies. In addition, ESPaDOnS data provide high-resolution line profiles, allowing detailed studies of the circumstellar environment, photospheric properties and structures, chemical abundances and stellar rotation.

Fig. 3 illustrates an ESPaDOnS spectrum of somewhat less than typical quality (peak S/N of 180:1) corresponding to the HAeBe star BF Ori, a UX~Ori-type star reported by Hillenbrand et al. (1992) to have $T_{\rm eff}=6750$~K, and by Grinin et al. (2001) to have have $T_{\rm eff}=8750$~K. In the left-hand frame of Fig. 3, we show the H$\alpha$ region of the spectrum. H$\alpha$ displays strong, double-peaked emission, and He~{\sc i}~$\lambda 6678$ is clearly visible in absorption. The detection of this absorption line is supported by identification of He~{\sc i}~$\lambda 5876$, also in strong absorption (also reported by Wade et al. 2006b). The appearance of these features in the spectrum of this star suggest that the published photospheric temperatures are seriously underestimated. Also, at high resolution many spectral lines resolve into complex profiles, which appear to be composed of two rotationally-broadened stellar components (separated by approximately 140~\kms), upon which are superimposed sharp, possibly circumstellar absorptions (e.g. Na~{\sc i}~D in Fig. 3, left panel inset, and Fe~{\sc ii}~$\lambda 5018$ in Fig. 3, right panel). Clearly the new spectra are rich in spectroscopic information, in addition to their unique value for diagnosing the magnetic field.

In Fig. 3 (right panel), we show the Stokes $V$ circular polarisation across the Fe~{\sc ii}~$\lambda 5018$ line of BF~Ori. No Stokes $V$ signature is detected in this line (or in any of its individual absorption components, for that matter), or in other absorption or emission lines of this star, notwithstanding the tentative detection of a magnetic field by Wade et al. (2006b). 

The ESPaDOnS observations are further analysed using the Least-Squares Deconvolution (LSD; Donati et al. 1997) multi-line procedure, extracting polarisation information from hundreds or thousands of lines in the observed spectrum to obtain the magnetic field diagnosis. The resultant high-S/N mean profiles allow the diagnosis of the magnetic field directly from the presence (or absence) of resolved Stokes $V$ signatures (see Fig. 4).  The longitudinal field can also be diagnosed: it is inferred from the first moment of the Stokes $V$ LSD profile (e.g. Wade et al. 2000b). However, because ESPaDOnS is a high-resolution spectropolarimeter, the inferred errors are very sensitive to the line profile characteristics - in particular $v\sin i$ and emission contamination. This results in a substantially broader distribution of longitudinal field uncertainties, the largest of which (for stars with the highest $v\sin i$ or lines most contaminated by emission) reach several hundred gauss. However, for the majority of the sample, the longitudinal field uncertainties are competitive with those obtained using FORS1, and for those stars which are most suitable spectroscopically, the longitudinal field uncertainties are substantially better than those from FORS1. For example, the LSD profiles of BF Ori yield a longitudinal magnetic field of $25\pm 32$~G, and those reported by Catala et al. (2006) have uncertainties as small as 10~G. The longitudinal field precision and accuracy obtainable with ESPaDOnS are illustrated in Fig. 4.

\begin{figure*}
\centering
\includegraphics[width=4.5cm,angle=-90]{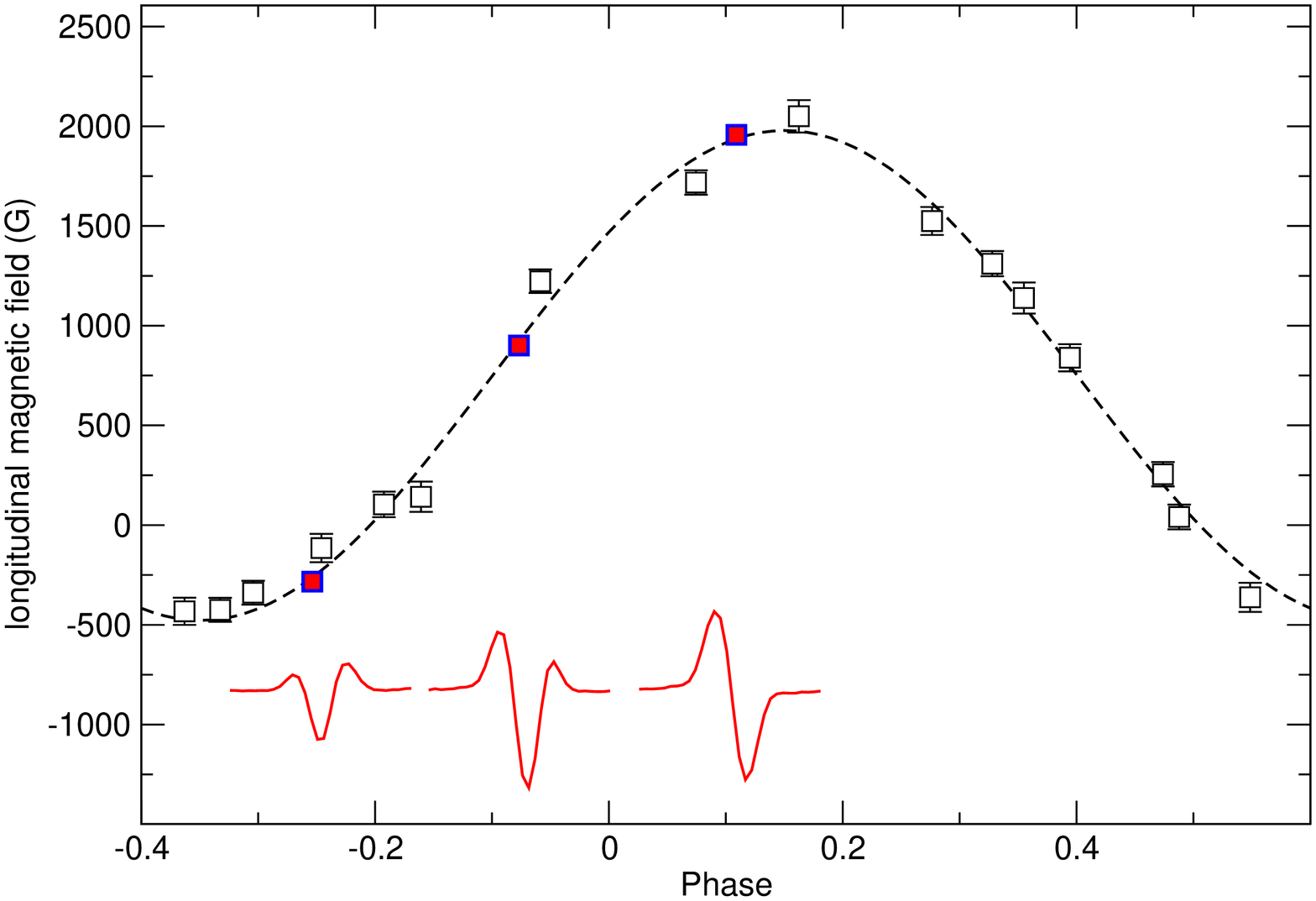}\includegraphics[width=4.5cm,angle=-90]{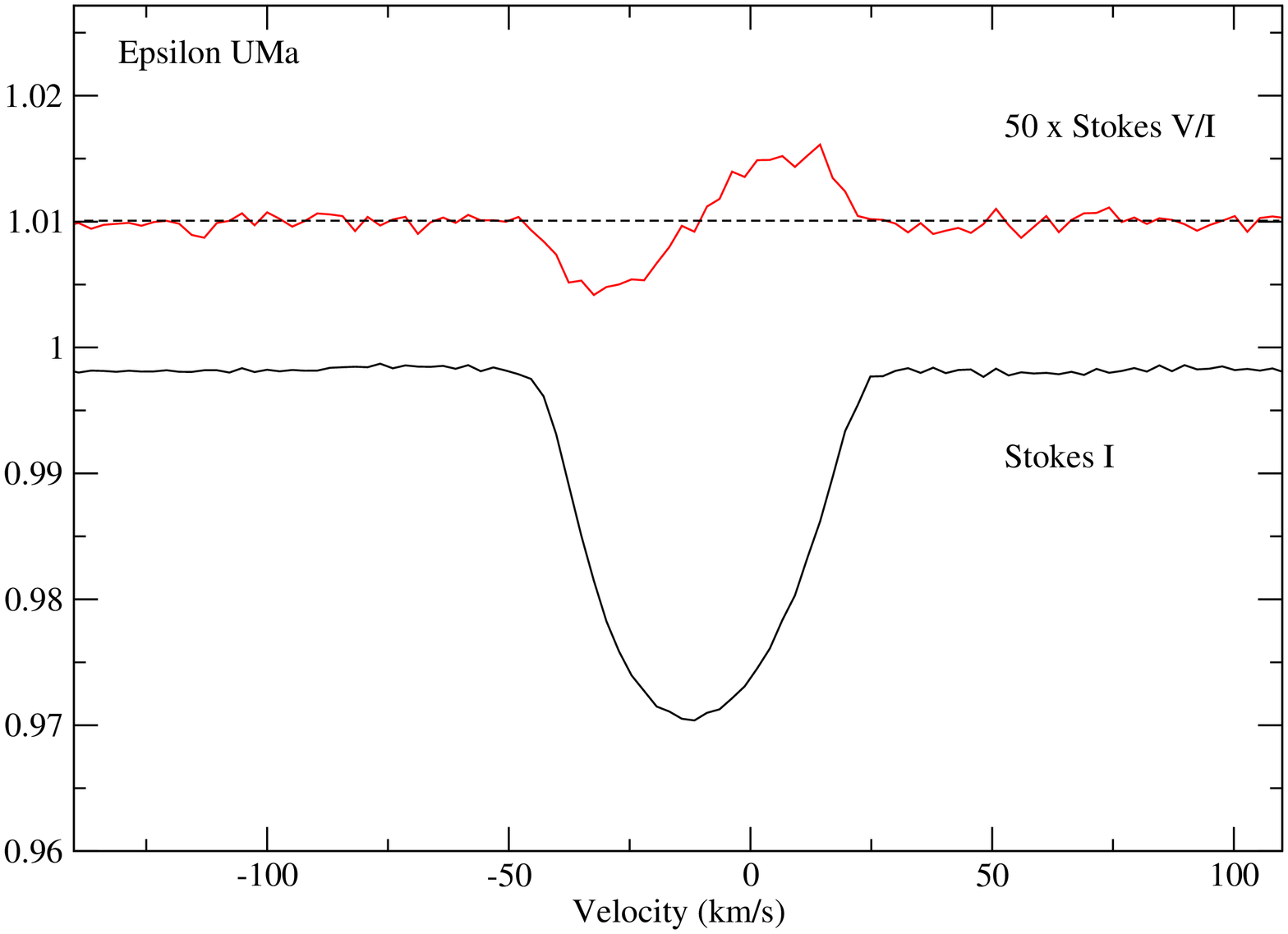}\includegraphics[width=4.5cm,angle=-90]{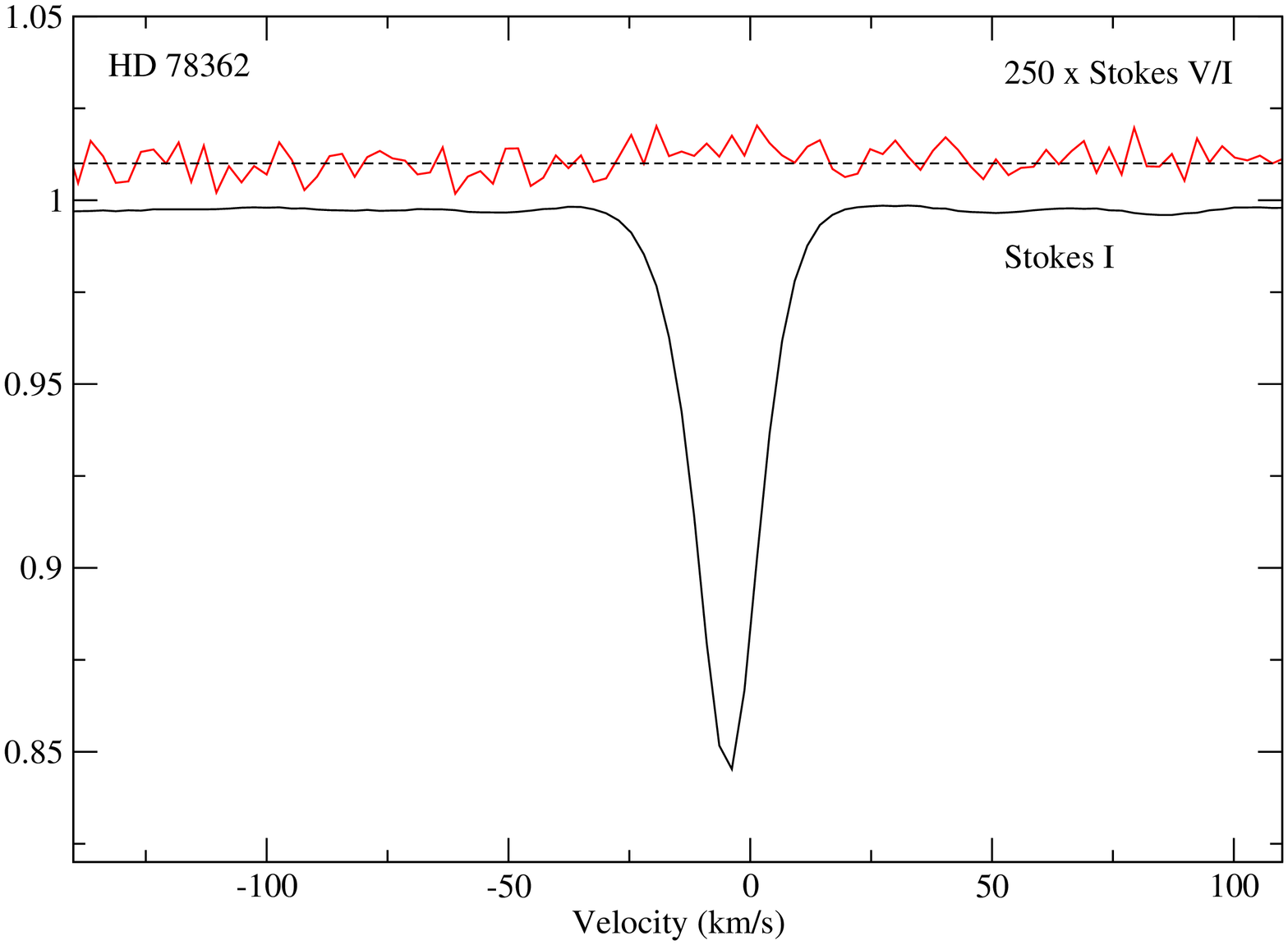}
\caption{{\em Left -}\ ESPaDOnS verification observations for magnetic and non-magnetic standard stars. {\em Left frame -}\ Longitudinal magnetic field variation of the cool magnetic Ap star HD~12098 obtained with the ESPaDOnS predecessor MuSiCoS (open symbols; Ryabchikova et al. 2005). Superimposed are 3 ESPaDOnS measurements (filled symbols) with error bars of about 25~G (also shown are their corresponding LSD Stokes $V$ profiles). Each ESPaDOnS measurement agrees with the MuSiCoS field variation to within 1$\sigma$. {\em Centre frame -}\ ESPaDOnS Stokes $I$ and $V$ LSD profiles of the weak-field Ap star $\epsilon$~UMa. The LSD profiles, detected with about $16\sigma$ confidence, correspond to a longitudinal magnetic field of $97\pm 6$~G. {\em Right frame -}\  ESPaDOnS Stokes $I$ and $V$ LSD profiles of the non-magnetic Am star HD~78362. The LSD profiles correspond to a longitudinal magnetic field of $0\pm 1$~G.}
\end{figure*}

Similar null magnetic field results are obtained for the majority of stars observed, including HD 144432, HD 31648 and HD 139614, in which Hubrig et al. (2004, 2006a) have claimed marginal magnetic field detections using FORS1, and in HD~36112, in which FORS1 data were reported to be ``suggestive'' of magnetic fields (Wade et al. 2006b). Nor do the ESPaDOnS spectra yield evidence for the systematic presence of polarisation signatures in lines diagnostic of the circumstellar environment, such as Ca~{\sc ii} H and K, Mg~{\sc ii} $\lambda 4481$, the H$\alpha$ emission line, the O~{\sc i} $\lambda\lambda$6300, 6364 emission lines, the IR calcium triplet, etc.  

On the other hand, magnetic fields are clearly detected in photospheric metal lines of 4 stars: HD~190073 (Catala et al. 2006), HD~72106 (Folsom et al., these proceedings), and HD~200775 and V380 Ori (Alecian et al., these proceedings). All of these stars have been detected in multiple observations during different observing runs, and two of these stars have been shown to exhibit coherent, periodic variations of their Stokes $V$ profiles (Alecian et al., these proceedings, Folsom et al., these proceedings). LSD profiles illustrating the detection of fields in these stars are shown in Figs. 5 and 6.

\begin{figure*}
\centering
\includegraphics[width=8cm]{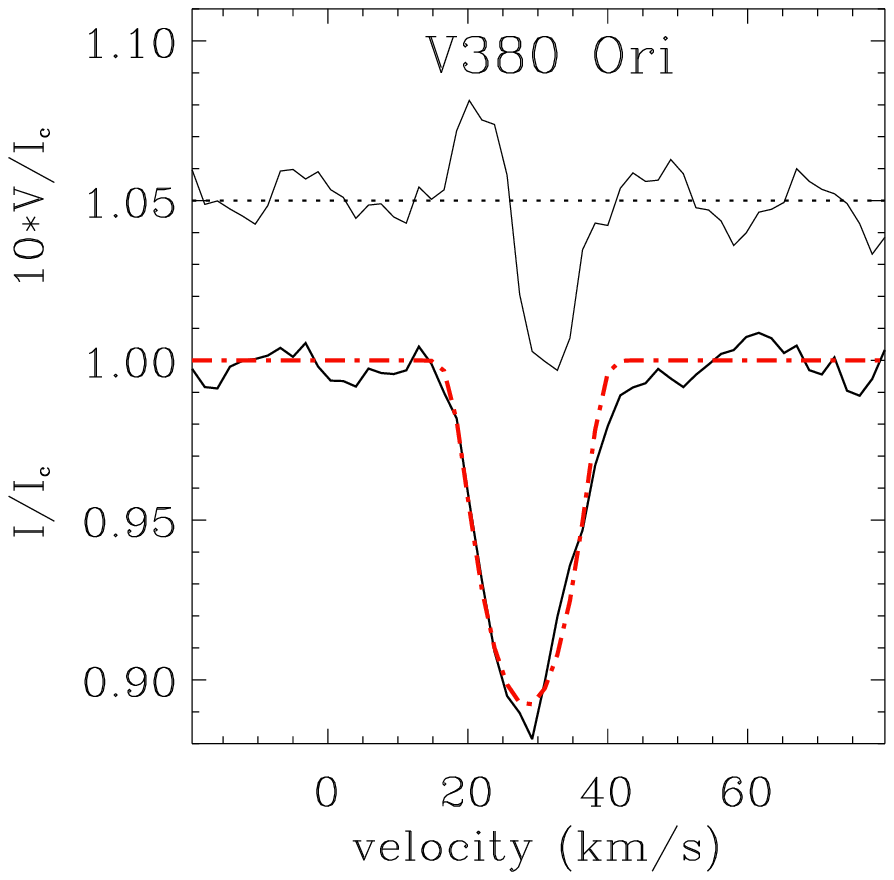}\includegraphics[width=8cm]{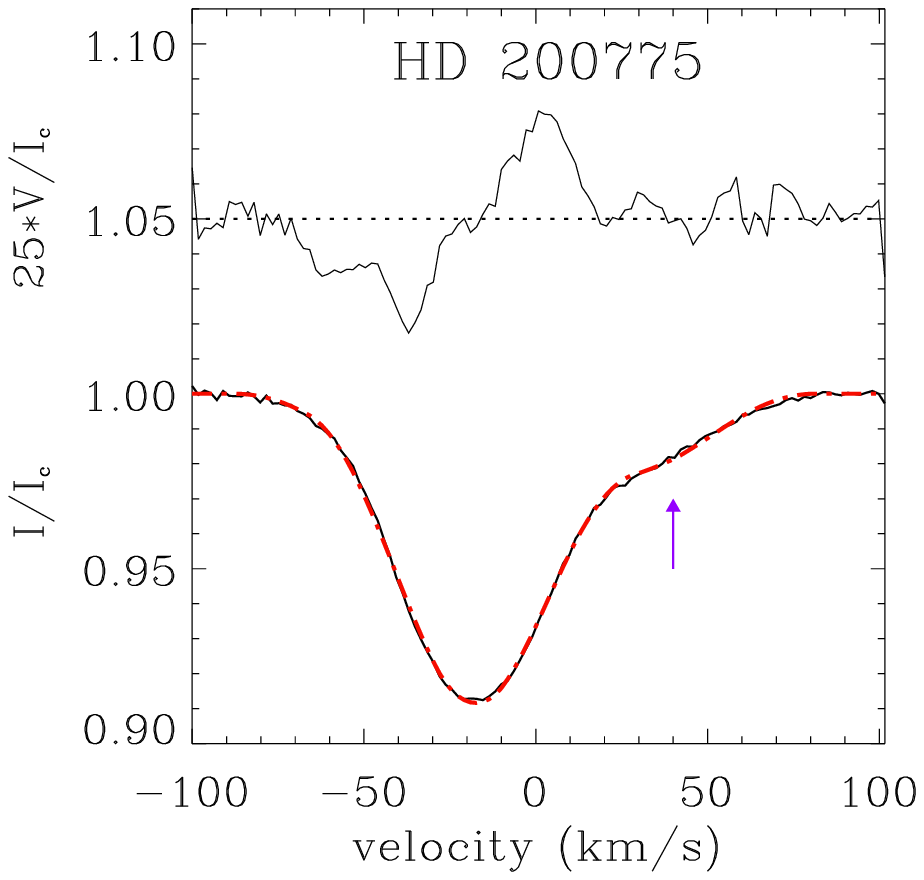}
\caption{ESPaDOnS Stokes $I$ and $V$ LSD profiles of V380~Ori (Left) and HD~200775 (Right). Strong, ordered magnetic fields are detected in both of these stars within the context of the ESPaDOnS survey. Modeling of the magnetic field geometries of these stars is presented by Alecian et al. (these proceedings).}
\end{figure*}

\begin{figure*}
\centering
\includegraphics[width=8cm,angle=-90]{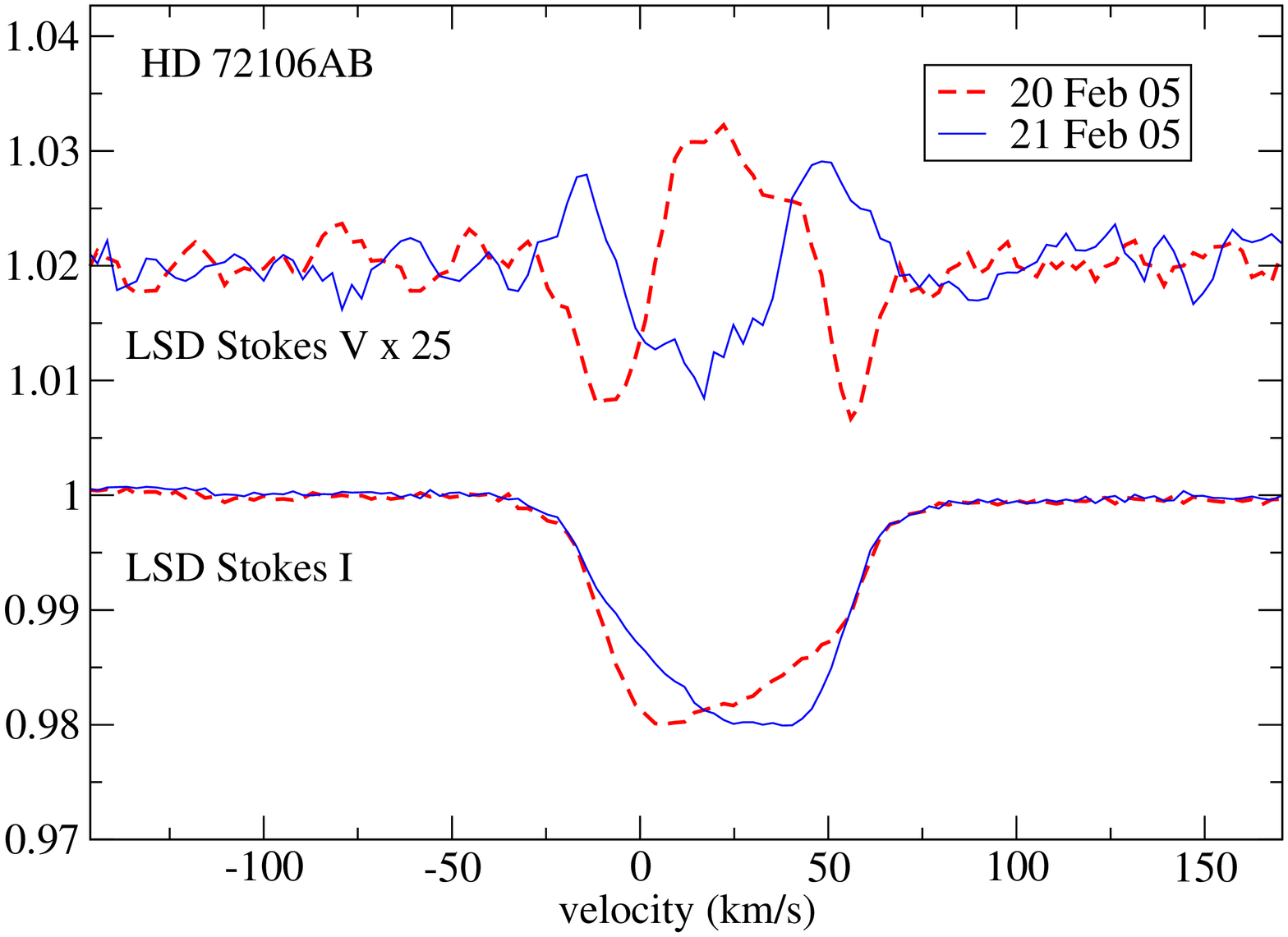}\hspace{0.25cm}\includegraphics[width=8cm,angle=-90]{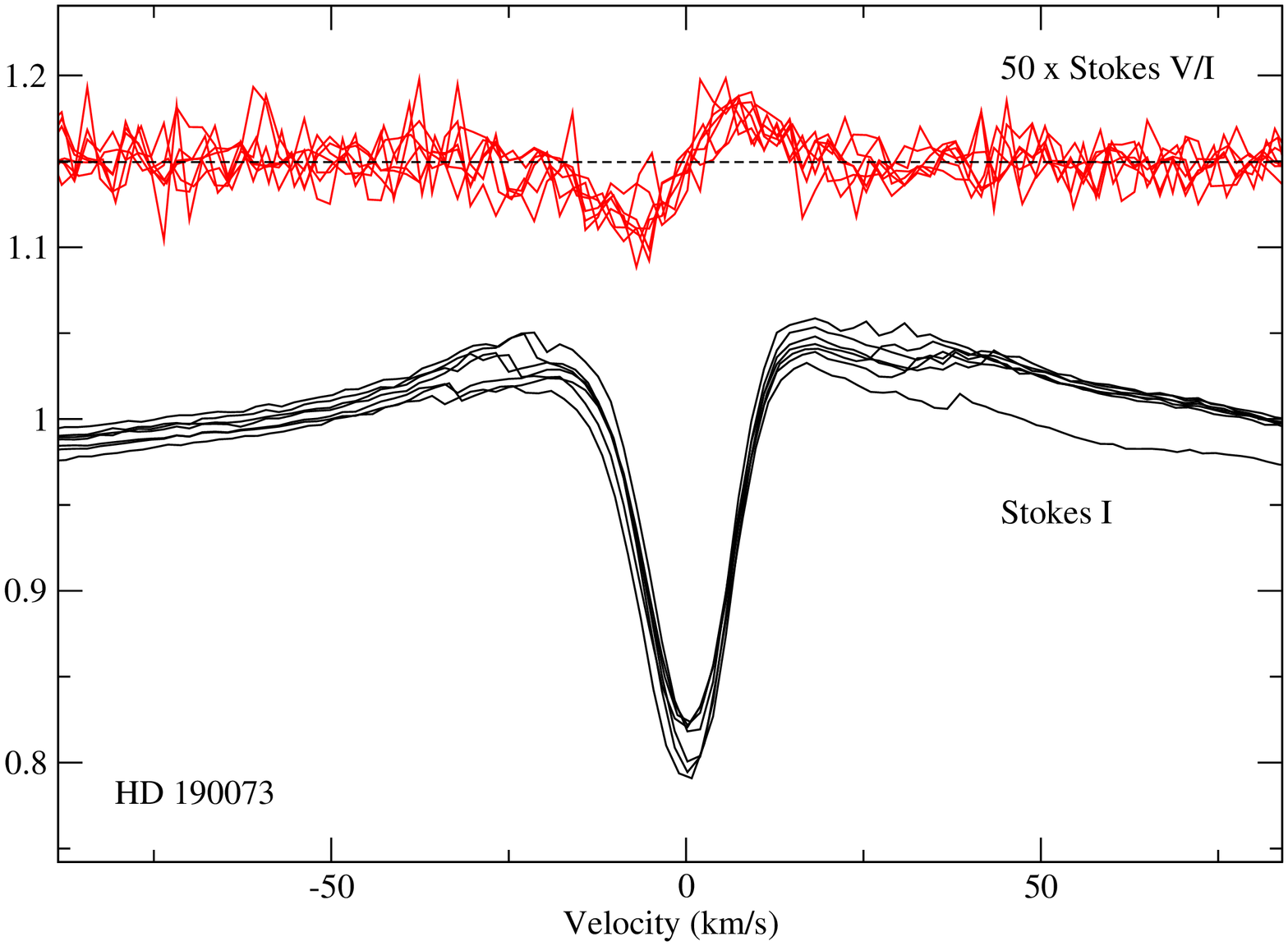}
\caption{ESPaDOnS Stokes $I$ and $V$ LSD profiles of HD~72106 (Left) and HD~190073 (Right). Strong, ordered magnetic fields are detected in both of these stars within the context of the ESPaDOnS survey. Further discussion of results for these stars is reported by Folsom et al. (these proceedings) and Catala et al. (2006).}
\end{figure*}

Although Hubrig et al. (2006b) suggest that LSD is inappropriate for magnetic field diagnosis of HAeBe stars, the results reported by Alecian et al. (these proceedings), Folsom et al. (these proceedings) and Catala et al. (2006) demonstrate that LSD can be used effectively for field diagnosis in HAeBe stars, even those stars with rather complex spectra. This conclusion is supported by the analysis of the large body of data accumulated within the context of the ESPaDOnS survey.

A final remarkable result comes from noting that 3 of the 4 stars detected in the ESPaDOnS survey (HD 72106, HD 190073 and V380 Ori) were also observed by Wade et al. (2006b), with no magnetic field detected. This is certainly partly due to the relative weakness of the longitudinal fields, but it also results from the inability of FORS1 to resolve line profiles into their (polarised) absorption components and (unpolarised) emission components, and to thereby isolate and detect the polarisation. This illustrates a clear advantage of high-resolution spectropolarimetry for studies of magnetism in HAeBe stars.

\section{General magnetic characteristics of Herbig Ae/Be stars}

For the large majority of stars studied by Wade et al. (2006b) and in the ESPaDOnS survey, no magnetic fields are detected. Wade et al. (2006b) report the results of Monte Carlo simulations aimed at modeling their FORS1 measurements. They compute synthetic distributions of longitudinal field measurements assuming populations of stars with various dipolar magnetic field characteristics. Comparing the observed and computed distributions (excluding measurements of HD~101412 and BF~Ori), they conclude that their observations have the following properties: they are consistent within statistical uncertainty with a distribution of non-magnetic stars; they are inconsistent with a uniform population of magnetic stars with aligned magnetic fields, if their dipole intensities $B_{\rm d}\gtrsim 300$~G; and they are inconsistent with a uniform population of magnetic stars with perpendicular magnetic fields, if $B_{\rm d}\gtrsim 500$~G. In addition to the large sample of stars in which magnetic fields are not observed, they have identified a small number of stars in which strong ($\sim 1$~kG) magnetic fields are detected, which must be organised on large scales (e.g. quasi-dipolar). 

 Although no similar analysis of the ESPaDOnS data has yet been performed, it is clear that the ESPaDOnS measurements are qualitatively consistent with all of the general conclusions of Wade et al. (2006b). In particular, the ESPaDOnS data confirm that magnetic fields are not detected in most stars studied, but that strong, ordered magnetic fields exist in a small fraction of HAeBe stars. In addition, a few apparently non-magnetic stars in the ESPaDOnS survey have been observed several times. This may allow much firmer upper limits to be placed on any magnetic fields that remain undetected in these stars. Finally, unlike the low-resolution FORS1 data, the ESPaDOnS profiles are sensitive to rather complex magnetic topologies (see, e.g., Petit et al. 2005). The absence of detectable circular polarisation in the high-resolution LSD profiles therefore also allows the general exclusion of a variety of more exotic field geometries. 

\section{Magnetic properties of HAeBe stars vs. main sequence A and B stars: testing the fossil-field hypothesis}

 {The primordial fossil-field hypothesis, which proposes that the magnetic fields of Ap/Bp stars are the slowly-decaying remnants of interstellar magnetic field swept up during star formation, requires that the magnetic properties of HAeBe stars be qualitatively the same as those of the main sequence A and B stars. It is therefore instructive to compare the known magnetic properties of the main sequence A and B type stars with the observed properties of HAeBe stars. In brief:

\begin{itemize}

\item Wade et al. (2006b) report that Monte Carlo simulations demonstrate that their complete sample of FORS1 observations of HAeBe stars is consistent with a magnetic field distribution similar to that of the main sequence A and B type stars. 

\item The total number of confirmed magnetic stars in the sample of Wade et al. (2006b) is 4, and the number of suspected magnetic stars is 1 (if we exclude BF Ori based on the results of the ESPaDOnS survey). This corresponds to a bulk incidence of (detected) magnetic stars of between 8\% and 10\% (taking into account the possibility that HD 101412 is not magnetic). This range is fully consistent with the canonical incidence of 5-10\% (e.g. Wolff 1968) of magnetic Ap and Bp stars on the main sequence.

\item The results reported by Wade et al. (2005), {Folsom et al. (these proceedings) and Alecian et al. (these proceedings) for HD~72106A, V380 Ori and HD~200775, and by Catala et al. (2006) for HD~190073, show that the magnetic fields of all these stars are organised on large scales, and that the fields of V380 Ori, HD~72106A and HD~200775 have important dipolar components.} The strong longitudinal field measured by Wade et al. (2006b) for HD~101412 also suggests that that its field is organised on large scales. It therefore appears likely that most, and possibly all, of the detected stars have magnetic fields that are structured on global scales, with important dipole components.

\item The mean rms longitudinal field measured for the small number of magnetic HAeBe stars detected so far is roughly 200~G, with an intrinsic dispersion of perhaps 100~G. This is consistent with the value estimated for a complete magnitude-limited sample of Ap stars by Bohlender \& Landstreet (1990; 330~G with an intrinsic dispersion of about 170~G). Moreover, a weaker mean field strength of HAeBe stars is expected if fossil magnetic flux is conserved during stellar evolution (Wade et al. 2006b).

\end{itemize}

{All of these conclusions and observations show that the magnetic characteristics of HAeBe stars (incidence, topology, intensity) are analogous to those of the Ap/Bp stars. Although further investigations are clearly required, we view this as strong evidence for the general correctness of the primordial fossil field hypothesis.}

\section{Activity and magnetospheric accretion in Herbig Ae/Be stars}

Some investigators have interpreted the strong spectroscopic, polarimetric and photometric activity of HAeBe stars to indicate that these stars are magnetically active (e.g. Catala et al. 1986). The magnetic results described in this paper do not support the proposal the the activity of most HAeBe stars is of magnetic origin.  

{Some investigators have also suggested that HAeBe stars are the higher mass analogues of classical T Tauri stars (CTTS), and that all of the phenomena associated with CTTS are also operating in HAeBe stars, including magnetospheric accretion (see for example Muzerolle et al. 2004).  

Magnetospheric accretion requires the presence of strong, large-scale magnetic fields at the stellar surface, and in the case of CTTS the predicted field strengths range up to several kG for specific stars. Wade et al. (2006b) use their data to evaluate three magnetospheric accretion models (see Johns-Krull et al. 1999). Estimating typical values of the mass, radius, rotation period and accretion rate, they compute that a dipole magnetic field with an intensity of about 500~G is required for magnetospheric accretion according to the models of K\"onigl  (1991) and Shu et al. (1994), and of about 100~G for the model of Cameron \& Campbell (1993). Based on the results of their Monte Carlo simulations, they conclude that magnetospheric accretion according to the theories of K\"onigl (1991) and Shu et al. (1994) is not generally occurring in HAeBe stars. 

The ESPaDOnS data should allow this issue to be studied further. First, the firmer upper limits achievable for undetected stars with multiple, high-precision observations should allow testing of the Cameron \& Campbell model for several stars. In addition, characterisation of the magnetic strengths and geometries of the detected magnetic stars should allow a detailed comparison of their properties with the predictions of all three models, and perhaps with more sophisticated numerical simulations (e.g. Yalenina et al. 2006).

}

\section{Development of chemical peculiarities}

A characteristic observational feature of magnetic A and B type stars on the main sequences is their strong photospheric chemical peculiarity. Although the basic mechanism responsible for the production of this phenomenon is known, the roles of the various separation and mixing processes, and the influence of the magnetic field, are understood only schematically.

The new observations of HAeBe stars provide the potential for the study of these phenomena at their earliest stages, allowing us to investigate the conditions and timescales required for the development of chemical peculiarity. 

Although this aspect has not yet been investigated in much detail, the ESPaDOnS observations suggest that chemical peculiarity is detected in only one of the magnetic stars identified so far: HD~72106A. As is reported by Folsom et al. (these proceedings), this star shows both strong peculiarities analogous to those of the Ap/Bp stars, as well as line profile variability indicative of chemical abundance patches. 

\section{Angular momentum evolution of magnetic intermediate-mass stars}

As discussed earlier, magnetic A and B type stars on the main sequence rotate significantly more slowly than non-magnetic stars of the same spectral type. This suggests that the magnetic field plays an important role in the shedding of rotational angular momentum in the magnetic stars, probably at the pre-main sequence stage (e.g., see St\c{e}pie\'n 2000). 

We are beginning to explore this aspect in some detail. First, we are able to study the rotational characteristics of magnetic HAeBe stars, first by using their measured $v\sin i$, and more accurately using their measured rotational periods and their inferred radii. All detected stars in both surveys appear to be slow rotators - hence significant angular momentum already seems to have been shed by these magnetic stars, even at these relatively young ages. 

It could be argued that the low projected rotational velocities of the detected magnetic stars result from a selection effect, because most of these stars were detected using high-resolution spectropolarimetry which becomes less sensitive to magnetic fields as $v\sin i$ increases. We counter this argument in two ways: first, the FORS1 observations, which are essentially insensitive to $v\sin i$, do not detect fields in any rapidly-rotating stars. In fact, the only star in which a reliable field detection seems to have been obtained with FORS1 (HD~101412) has very low $v\sin i$ ($\sim 7$~\kms, Guimaraes et al. 2006), fully consistent with the ESPaDOnS result. Secondly, recent ESPaDOnS observations of HAeBe stars in young open clusters (aimed at studying the influence of environment and evolution of magnetism and rotation) demonstrate that magnetic fields can be detected in rapidly-rotating HAeBe stars (see Fig. 7).

\begin{figure*}
\centering
\includegraphics[width=8cm,angle=-90]{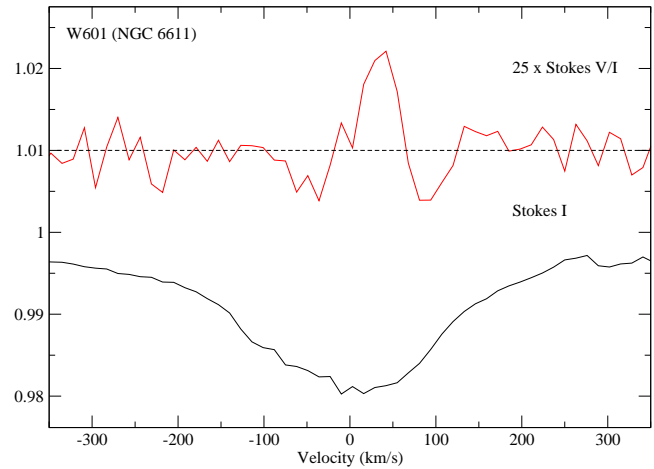}
\caption{ESPaDOnS Stokes $I$ and $V$ LSD profiles of the rapidly-rotating ($v\sin i\sim 180$~\kms) Herbig Be star W601 in NGC~6611, showing a definite detection of circular polarisation across the mean line. This detection demonstrates that magnetic fields are present in some HAeBe stars with high $v\sin i$, and that LSD is capable of detecting them.}
\end{figure*}

\section{Conclusion}

{In this paper we have reviewed the results of recent large-scale surveys of the magnetic properties of Herbig Ae/Be stars, demonstrating the clear detection of strong, organised magnetic fields in a small of the observed stars, and the apparent absence of magnetic fields in the large majority. We have reviewed the observed characteristics of these fields, and evaluated their consistency with the primordial fossil field hypothesis. Finally, we have explored the implications of these new data for our understanding of activity and magnetospheric accretion in PMS intermediate-mass stars, for the evolution of rotational angular momentum, and for the development of photospheric chemical peculiarity.

\begin{acknowledgements}
This research has been partially funded by the Natural Sciences and Engineering Research Council of Canada (NSERC), and the Canadian Department of National Defence Academic Research Programme (ARP).
\end{acknowledgements}

\end{document}